\documentstyle[12pt,openbib,epsf]{article}

\newcommand{\postscript}[2] {\setlength{\epsfxsize}{#2\hsize}
\centerline{\epsfbox{#1}}}
\begin{document}

\title{\bf  Universal scaling  in BCS superconductivity in two dimensions
in non-$s$ waves}
\author{Sadhan K. Adhikari\thanks{John Simon Guggenheim Memorial Foundation
Fellow}
 and Angsula Ghosh\\
Instituto de F\'{\i}sica Te\'orica, Universidade Estadual Paulista\\
01.405-900 S\~ao Paulo, S\~ao Paulo, Brazil\\}

\date{\today}

\maketitle

\begin{abstract}

The solutions of a renormalized BCS model are studied in two space dimensions
in $s$, $p$ and $d$ waves for finite-range  separable potentials. The gap
parameter, the critical temperature $T_c$, the coherence length $\xi$  and
the jump in specific heat  at $T_c$ as a function of zero-temperature
condensation energy exhibit universal scalings. In the weak-coupling limit,
the present model yields a small 
 $\xi$ and large $T_c$  appropriate
  to those for high-$T_c$ cuprates.
The specific heat, penetration depth and thermal conductivity as a function
of temperature show universal scaling in $p$ and $d$ waves.
 
{PACS number(s): 74.20.Fg, 74.72.-h, 74.76.-w}

\end{abstract} 


\section{Introduction}

At low temperature and 
in the weak coupling  limit,  a collection of weakly interacting electrons,
spontaneously form  large overlapping Cooper pairs \cite{e} leading to the
microscopic Bardeen-Cooper-Schreiffer (BCS) theory of superconductivity
\cite{8,t}. For usual superconductors, the $s$-wave BCS theory yields
$\xi k_F \sim 1000$ in agreement with experiments, where $\xi$ is the
coherence length and $k_F$ the Fermi momentum.  There has been renewed
interest in this problem with the discovery of high-$T_c$
superconductors.  The  high-$T_c$ materials  have a small  $\xi$:
  $\xi k_F \sim 10$ \cite{c,6}.  Inspite of much effort, the normal state
of  the high-$T_c$ superconductors has not been satisfactorily
understood \cite{n}.  There are  controversies about the appropriate
microscopic hamiltonian, pairing mechanism, and gap parameter \cite{c,6,d}.

Many high-$T_c $ superconductors have a conducting structure similar to a
two-dimensional layer of carriers \cite{c,d,u}, which suggests the use of
two-dimensional  models.
Moreover, there are evidences  that  the high-$T_c$ 
cuprates have singlet $d$-wave Cooper pairs  and the gap parameter
has $d_{x^2-y^2}$ symmetry in  two dimensions \cite{d}.  Recent
measurements of penetration depth $\lambda(T)$  \cite{h} and superconducting
specific heat at different temperatures $T$ \cite{s} and related theoretical
analyses \cite{t1,t2} also support this point of view.  According to the
isotropic $s$-wave BCS theory,  as $T\to 0$, both observables exhibit
exponential dependence on  $T$ \cite{t,t2}.  The experimental power-law
dependence of these observables on $T$ can be explained  by
considering anisotropic gap parameter with node(s) on the Fermi surface in
higher partial wave \cite{t1,t2}. It seems that some  properties of the
high-$T_c$ cuprates can be  explained by the $d$-wave
BCS equation in two dimensions for weak coupling.
Yet there has been no systematic study of the BCS equation in non-$s$ waves
in two dimensions.

The  BCS theory 
considers $N$ electrons of spacing $L$, interacting via a weak potential of 
short range $r_0 $ such that $r_0 << L$ and $r_0 << R$ where $R$ is the
pair radius.  When suitably scaled, most  properties of the system should be
insensitive to the details of the potential and be universal functions of
the dimensionless variable $L/R$ \cite{L}. 
The usual BCS treatment \cite{8} employs a 
phonon-induced two-electron potential of moderate range. 
In this paper we study the weak-coupling BCS problem in two dimensions 
 for $s$, $p$, and
$d$  waves    
with two objectives in mind. The first objective is to identify 
the universal nature of the solution and its relation to   
to high-$T_c$ superconductors.       
 The second objective
is to find out to what extent the universal nature of the solution
  is modified in the presence of
realistic finite-range (nonlocal separable) potentials.  Instead of solving 
the BCS equation on the lattice with appropriate symmetry, we solved the 
equations in the continuum. This procedure should suffice for  present
objectives. 

In place of the phonon-induced BCS model we employ a renormalized BCS model 
with separable potential which has certain advantage. 
The renormalized BCS 
model leads to convergent result even in the absence of potential form
factors or momentum/energy cut off, as required in the standard BCS model. 
The original BCS model 
yields a linear correlation between $T_c$ and $T_D$, where $T_D$ is the Debye
temperature. This correlation was fundamental in explaining the observed
isotope effect in conventional superconductors. The high-$T_c$ materials 
exhibit 
an anomalously negligible isotope effect and  a linear correlation
between $T_c$ and $T_F$, where $T_F$ is the Fermi temperature.
 This suggests a  different interaction  for
superconductivity in the high-$T_c$ materials.  The present renormalized
model yields a linear scaling between $T_c$ and $T_F$. Because of this
scaling with $T_F$, unlike in the phonon-induced BCS model, the present $T_c$
can be large and appropriate for the high-$T_c$ cuprates in the weak coupling
region.
The present  model 
also produces an appropriate $T_c/T_F$ ratio 
and a small $\xi k_F$ in the weak coupling  region in accord with recent 
experiments \cite{u} on high-$T_c$ cuprates. In spite of  these we are 
aware that there are controversies in the description of the high-$T_c$
materials, for example, the microscopic hamiltonian and the pairing mechanism.
Also, the normal state away from $T_c$ seems to be very different from a 
standard Fermi liquid. However, we find that there are certain
characteristics of these high-$T_c$ materials which can be studied 
within the present  renormalized mean-field BCS model based on the standard
 Fermi liquid theory.

Previously, there have been studies of this problem in two 
dimensions in terms of  two-body binding 
in vacuum and  Cooper pair binding  
employing  short- and  zero-range potentials \cite{c,6,us}.  Such studies
have not fully revealed the universal nature of the solution.  In the
present paper we employ the zero-temperature 
condensation energy per particle, $\Delta U$,
 of the BCS condensate  as the reference
variable for studying the BCS problem. As the condensation energy increases,
 one
passes from the weak to medium coupling. Also, $\Delta U$  is a
physical 
observable and is the appropriate reference variable as we shall see. 
In this paper  we
calculate  the zero-temperature gap parameter $\Delta(0)$,  the 
critical temperature $T_c$ and  the
 specific heat per particle $C_s(T_c)$  in all partial waves (denoted by  
 angular momentum $l$)  and
the zero-temperature pair size in $s$ wave. We find that these observables  
obey robust 
universal scaling as functions of $\Delta U$  valid over several 
decades 
in the weak to 
medium coupling region  independent of the potential 
 range parameter.

Here,
we also  calculate the temperature dependence of different quantities, such
as, the gap parameter $\Delta(T)$,    $C_s(T)$, $\lambda(T)$ and
thermal conductivity $K(T)$  
 for $T<T_c$.  Of these, the $T$
dependence of $C_s(T)$,  $\lambda(T)$ and $K(T)$ are specially interesting.
For isotropic $s$ wave,  the BCS theory yields exponential
dependence on temperature as $T \to 0$ for  
these observables  independent of space dimension \cite{t,t2}. 
 The observed power-law dependence of some these
observables can  be explained with anisotropic gap parameter in non-$s$
waves with node on the Fermi surface.  We find  universal power-law dependence
in both cases for non-$s$ waves independent of the range of potential.  For
$l\ne 0$ we find $C_s(T) \approx 2C_n(T_c) (T/T_c)^2$ 
and $K_s(T) \approx K_n(T) (T/T_c)^{1.2} $ valid for almost
the entire range of temperature.  
The suffix $n$ and $s$  refer to normal and superconducting
states, respectively. Similar dependence was predicted from an
analysis of experimental data \cite{s} as well as from a calculation based on
Eliashberg equation \cite{t1}.  In order to detect the  anisotropy in 
$\Delta(T)$
it is appropriate to  consider the
function $\Delta \lambda
\equiv (\lambda(T)-\lambda(0))/\lambda(0)$ \cite{t2}.
We find that for small $T$
it behaves as $\Delta \lambda \sim (T/T_c)^{1.3}$. 
Similar power-law dependence 
was conjectured before \cite{t2}.

From the weak-coupling BCS equation we establish the
following  relations analytically: $2\Delta(0)/T_c \approx $3.528 (3.026),
$\Delta(0)$ = $2\sqrt{\Delta U}$ ($2\sqrt{\Delta U}$), $T_c\approx$ 
1.134$\sqrt{\Delta U}$ ($1.322\sqrt{\Delta U}$), $\Delta
C/C_n(T_c)\approx $
 1.43 (1.05), $C_s(T_c)\approx$  9.065$\sqrt{\Delta U}$ (8.915$\sqrt{\Delta
U}$), ${\Delta U}/U_n(T_c) \approx$ 0.473 (0.348), for $l=0$ ($\ne 0$).  Here
  $\Delta C$  is the jump in specific heat
at $T=T_c$, $C$ ($U$)  is the specific heat (internal energy) per particle. 

The plan of the paper is as follows. In section 2 we present the renormalized
BCS model. In section 3 we present an account of our  numerical 
study. Finally, in section 4 we present a brief summary of our findings.

\section{Renormalized BCS model}

The  two-body problem, Cooper  and  BCS models all exhibit ultraviolet
divergences for zero range potential and require regularization or
renormalization to produce finite  results \cite{c}. The usual BCS model
employs a cut-off regularization (Debye cut off) for obtaining finite
observables.  In momentum space, the standard BCS model interaction 
is taken to be 
 constant   for momenta 
between $2m(E_F-E_D)/\hbar^2$ and $2m(E_F+E_D)/\hbar^2$ and
zero elsewhere with  $E_D$  ($E_F$) the Debye (Fermi) 
energy.  This implies a moderate
 range of the interaction.  This  potential is a physically
motivated phonon induced electron-electron one \cite{8}.
The present renormalized BCS equation may lead to well-defined
solution without requiring a cut off even in the absence of potential form
factors.  The present model with finite-range potential also leads to a
well-defined mathematical problem.

We consider a two-body system, each of mass $m$, in the center of mass  frame
\cite{c}. The single (two) particle energy is given by $\epsilon_q =\hbar^2
q^2/2m$  ($2\epsilon_q$), where $q$ is the wave number.
We consider 
a purely attractive short-range separable potential in partial wave 
$l$: 
\begin{equation}\label{pot}V_{ \bf pq}=-V_0 c_l 
g_{pl} g_{ql}\cos (l\theta)  
\end{equation}   where $\theta$ is the angle
between  vectors ${\bf p}$ and ${\bf q}$ and $c_l= 1$ (2) for $l=0$ 
$(\ne 0)$ as in Ref. \cite{a}. Here $g_{pl}$ and $g_{ql}$ are potential 
form factors and $V_0$ is the potential strength.
The potential $V_{ \bf pq}$ is the effective electron-electron potential 
in the superconductor in the presence of lattice and other electrons.  
The Schr\"odinger equation in this case leads to the following 
condition for a two-particle bound state in vacuum of  
binding  $B_2$ is 
\begin{equation} 
V_0^{-1}= c_l\sum _{\bf q} g_{ql}^2 \cos ^2(l\theta) (B_2+2\epsilon_q)^{-1}
\label{1}
\end{equation}
where  $\theta$ is the angle of vector ${\bf q}$
and $\epsilon_q =\hbar^2
q^2/2m$ with $m$  the mass. In order to simplify the
notation, the trivial $l$ dependence of many quantities, such as $V_{\bf pq}$,
$B_2$, etc., is suppressed.

For attractive potential (\ref{1}), two electrons on the top of the full 
Fermi sea at zero temperature shows pairing instability.
Singlet (triplet) pairing is assumed to take place  in even (odd)
partial waves. 
The Cooper pair problem for two electrons above the filled Fermi sea for
this potential is given  by \cite{e,8} $
V_0^{-1} =c_l \sum_{q> 1} g_{ql}^2\cos^2(l\theta)(2\epsilon_{q} -2\hat E)^{-1},$
with Cooper binding $B_c \equiv 2- 2\hat E$.  
Using  (\ref{1}), the
Cooper  problem is written as
\begin{eqnarray}
\sum_q g_{ql}^2\cos^2(l\theta)
(B_2+2\epsilon_q)^ {-1}& - & \sum_{q> 1} g_{ql}^2\cos^2(l\theta)
(2\epsilon_q-2\hat E)^ {-1} \nonumber \\ &= &0.\label{126}
\end{eqnarray}

At a finite $T$, one has  the following BCS  and number
equations 
\begin{eqnarray}
\Delta_{\bf p}& =& -\sum_{\bf q} V_{\bf pq}\frac{\Delta_{ \bf q}}{2E_{\bf
q}}\tanh
\frac{E_{\bf q} }{2T}, \label{130}\\ N &=& \sum_{\bf q}
\left[1-\frac{\epsilon_q-\mu}{E_{\bf q}}
\tanh\frac{E_{\bf q}}{2T} \right] \label{140} \end{eqnarray}
 with $E_{\bf q} = [(\epsilon_q - \mu )^2 + |\Delta_{\bf q}|^{2}]^ {1/2},$
where $\Delta_{\bf q}$ is the gap function. Unless the units of the variables
are explicitly mentioned,
in  (\ref{1}) $-$ (\ref{140})
and in the following all energy/momentum  variables are expressed in units of
$E_F$, such that $\mu \equiv \mu/E_F$, $T \equiv T/T_F$, $q\equiv q/k_F$,
$E_{\bf q}\equiv
E_{\bf
q}/E_F$, etc, where $\mu$ is the chemical potential. Here  $\Delta _{\bf q}$
has the 
following anisotropic form: $\Delta _{\bf q} \equiv 
g_{ql} \Delta_0 \sqrt c_l \cos(l \theta)$ where 
 $\Delta_0$ and $g_{ql}$  are dimensionless.  The usual BCS gap is defined 
 by  $\Delta(T)=g_{q(=1)l}\Delta_0$, which is the root-mean-square average of
   $\Delta _{\bf q}$ on the Fermi surface. Using the above form of
$\Delta_{\bf q}$, the BCS equation (\ref{130})  can be written as
\begin{equation}\label{new}
\frac{1}{V_0}=c_l\sum_{\bf q}g_{ql}^2\cos^2(l\theta)
\frac{1}{2E_{\bf q}}\tanh \frac{E_{\bf q}}{2T}. 
\end{equation}
Now     (\ref{1}) and
(\ref{new}) lead to
\begin{eqnarray}
\sum_{\bf q}g_{ql}^2\cos^2(l\theta) \biggr[\frac{2}{2\epsilon_q+B_2}-
\frac{1}{E_{\bf q}}\tanh \frac{E_{\bf q}}{2T} \biggr]=0.
\label{160} \end{eqnarray}
The summation is evaluated according to 
\begin{equation} \sum _{\bf q} \to\frac {N}{4\pi} 
\int d\epsilon_ q  d\theta 
\equiv \frac 
{N}{4\pi} 
\int_0^\infty d\epsilon_ q \int_0^{2\pi}d\theta
\label{n1}\end{equation}
where $N$ is the number of electrons. As in the standard BCS model,
we have a constant density of state in (\ref{n1}) but now with the 
generalization to include the angular dependence. With the help of
(\ref{n1}),
(\ref{140}) and (\ref{160}) can be explicitly written as
\begin{equation}
\int d\epsilon_ q  d\theta 
\biggr[1-\frac{\epsilon_q-\mu}{E_{\bf q}}\tanh
\frac{E_{\bf q}}{2T}\biggr] ={4}\pi \label{16}\end{equation}
\begin{eqnarray}
\int  d\epsilon_ q  d\theta 
g_{ql}^2\cos^2(l\theta) \biggr[\frac{2}{2\epsilon_q+B_2}-
\frac{1}{E_{\bf q}}\tanh \frac{E_{\bf q}}{2T} \biggr]=0\label{15}
\end{eqnarray}
respectively. In terms of Cooper-pair binding Eq. (\ref{15})  can be
rewritten as 
\begin{eqnarray}
\int_0^{2\pi}    d\theta 
\cos^2(l\theta) &. & \left[ \int_1^\infty d\epsilon_ q
\frac{g_{ql}^2}{\epsilon_q -  \hat E}-\int_0^\infty
d\epsilon_ q
\frac{g_{ql}^2}{E_{\bf q}}\tanh \frac{E_{\bf q}}{2T} \right]\nonumber \\
& = & 0\label{15q}
\end{eqnarray}
 Even in the absence of potential form factors $g_{pl}=1$,
(\ref{16}), (\ref{15}) and (\ref{15q}) 
 are well defined without any energy/momentum
cut-off, though each part of the integral in these equations
diverges separately. This is why this model is termed renormalized.
Potential (\ref{pot}) with $g_{pl}=1$ is the zero-range 
delta-function potential.  The standard BCS model uses essentially
the above potential with energy/momentum cut-off for obtaining convergence. 
In renormalized  equation (\ref{15})
the usual energy/momentum cut-off and the potential strength $V_0$
have been eliminated in favor of the two-body binding $B_2$. 
Recently, the role of renormalization in nonrelativistic quantum 
mechanics has been discussed \cite{re}.

Equation (\ref{15q}) has following analytic solutions for 
$s$-wave zero-range potential ($g_{q0}=1$) in the weak coupling 
 limit ($\mu = 1$). At
$T=0$, $ \Delta(0) = \sqrt {2 B_c }$. At $T=T_c$ ($\Delta=0$),  we have:
${T_c} = {\exp(\gamma)}\sqrt{2 B_c}/\pi 
\approx 0.8\sqrt{B_c}$ where $\gamma = 0.57722..$.
The standard BCS model yields in this case $T_c/T_D = 
{\exp(\gamma)}\sqrt{2 B_c}/\pi\sqrt{T_D}
\approx 0.8\sqrt{B_c/T_D}$ where $T_D$ is the Debye 
temperature. To illustrate the advantage of the renormalized model let us
consider a specific example with $T_D =300$ K, $T_F= 3000$ K and $B_c =10$ K.
We take 
$B_c < 10$ K as defining the weak-coupling region. With  $B_c=10$ K, the 
standard BCS model yields $T_c = 44$ K, whereas the present renormalized
model yields
$T_c= 138$ K. Hence for same coupling, the renormalized model leads to 
an enhanced $T_c$.   
In the standard BCS  model one has to have a 
much larger $B_c$, clearly outside the weak-coupling region, in order to 
have $T_c >$ 100K. 

  The universal ratio $ 2\Delta(0)/T_c = 2\pi /\exp(\gamma)\approx
3.528$  remains unchanged for $s$ wave in three dimensions as
well as for the trivial case of anisotropic pairing ($\sim
\exp(il\theta)$) in two dimensions for $l\ne 0$ \cite{us}.  In 
Ref. \cite{us} we essentially changed the potential form-factors without
introducing any explicit angular dependence in the BCS and  number
equations and find that the universal nature of the solution was unchanged
with such change. Hence we expect to extract certain universal properties
of (\ref{15}) analytically for $l\ne 0$ by employing
the correct angular distribution and 
no potential form factors ($g_{ql}=1$). 

Next we study  (\ref{15}) analytically 
for  weak coupling ($\mu=1, \Delta << 1$) and for $g_{ql}=1$.
At $T=0$,  (\ref{15}) can be integrated  to yield
$$
\int_0^{2\pi} d\theta \cos^2(l\theta)\ln \frac
{[1+\Delta^2(0)c_l\cos^2(l\theta)]^
{1/2} -1}{B_2}=0,
$$
which 
for $\Delta(0) << 1$ and $l\ne 0$  reduces to
$$
\ln \frac{c_l\Delta^2(0)}{2B_2}= 
-\frac{1}{\pi}\int_0^{2\pi}d\theta \cos^2(l\theta)
\ln\cos^2(l\theta)\equiv 0.3863,$$
or,  $\Delta(0)\approx 1.213\sqrt{B_2}.$
At $T=T_c$,  we again have
${T_c} = {\exp(\gamma)}\sqrt{2 B_2}/\pi $, so that 
we have  the following new universal
constant for $l\ne 0$:  $2\Delta(0)/T_c \approx 3.026$. 

The condensation energy per particle at $T=0$  is defined by \cite{t}
$$
\Delta U\equiv {|U_s-U_n|}=\frac{1}{N}\sum_{{\bf q}
(q<1)}2\zeta_q-\frac{1}{N}\sum_{\bf q}(\zeta_{q}
-\frac{\zeta_q^2}{E_{\bf q}}-\frac{\Delta_{\bf q}^2}{2E_{\bf q}}) $$
where $\zeta_q=(\epsilon_q-\mu)$. 
This can be evaluated straightforwardly to lead to \cite{t}
$$
\Delta U = \frac{1}{8\pi}\int_0^{2\pi}c_l 
\Delta^2(0) \cos^2(l\theta) d\theta
$$
which yields $\Delta(0)  = 2\sqrt{\Delta U}$ for all $l$.  Using the
universal relation between $\Delta(0)$ and $T_c$ given above one has $T_c
\approx 1.134\sqrt {\Delta U}$ $(1.322\sqrt {\Delta U})$ for $l=0$ $(l \ne
0)$. For all $l$,  $U_n(T_c)= \pi ^2 T_c^2/6$, so that $\Delta U/U_n(T_c)
\approx $0.473 (0.348) for $l=0$ $(\ne 0)$.

The superconducting specific heat per particle  is given by
\begin{equation}
C_{s}= \frac{2}{NT^2}\sum_{\bf q}   f_{\bf q}(1-f_{\bf q})
\left( E_{\bf q}^2-\frac{1}{2}T\frac{d\Delta_{\bf q}^2}{dT} \right)\label{sp} 
\end{equation} 
where $f_{\bf q}=1/(1+\exp( E_{\bf q}/ T))$.
The normal specific heat $C_n$ is given by  (\ref{sp}) with $\Delta_{\bf q}
=0.$
The jump in specific heat  per particle 
at $T=T_c$ ($\Delta(T_c)=0$), $\Delta C \equiv  [C_s-C_n]_{T_c}$
 is given by
\cite{t}
\begin{eqnarray}
\Delta C =-\frac{1}{NT_c} \sum_{\bf q}\left[f_{\bf q}(1-f_{\bf q})
\frac{d\Delta_{\bf q}^2}{dT}
\right]_{T_c}. 
\label{dc}\end{eqnarray}
In the special case $g_{ql}=1$, the radial integral in  (\ref{dc})
can be evaluated as in Ref. \cite{t} and we get 
\begin{eqnarray}\Delta C=-
\int c_l
\frac{d\epsilon_q d\theta}{4\pi T_c}\biggr[f_{\bf q}(1-f_{\bf q})
 \frac{ d\Delta^2(T)}{dT}\biggr]_{T_c}\cos^2 (l\theta) .   \end{eqnarray}
 This
 leads to \cite{t}
  $\Delta C= -(1/2)[d\Delta^2(T)/dT]_{T=T_c}$ for  all $l$.
In a systematic (numerical) study we find  $\Delta(T)/\Delta(0) =
B(1-T/T_c)^{1/2}$ valid for $T\approx T_c$, with $B\approx 1.74$ (1.70)
for $l =0$ ($\ne 0$). Using the   value of $B$ 
 and the universal ratio $2\Delta(0)/T_c$, we obtain
$\Delta C/C_n(T_c)
 \approx$ 1.43 (1.00) for $l=0$ $(\ne 0),$ where 
 $C_n(T)=\pi^2T/3 $. Consequently,
$C_s(T_c)/\sqrt{\Delta U}\approx 9.065 $ 
$( 8.915)$  and $T_cC_n(T_c)/\Delta U 
\approx  4.229$  (5.749)  for $l=0$ $(\ne 0)$.

The penetration depth $\lambda$
is defined by \cite{t}
\begin{equation}
\lambda^{-2}(T)=\lambda^{-2}(0)\left[ 1-\frac{2}{NT} 
\sum_{\bf q}f_{\bf q}(1-f_{\bf q})\right].
\end{equation}

The thermal conductivity ratio $K_s(T)/K_n(T)$  is defined by \cite{con}
\begin{equation}
\frac {K_s(T)}{K_n(T)} =\frac  {\sum_{\bf q}\zeta_q E_{\bf q}
f_{\bf q}(1-f_{\bf q}) 
}{\sum_{\bf q}\zeta_q^2
f_{\bf q}(1-f_{\bf q}) 
}.\label{con}
\end{equation}
The denominator of (\ref{con}) essentially corresponds to the normal-state 
thermal conductivity with the BCS gap set equal to zero. 

The dimensionless $s$-wave pair radius
 defined by $\xi ^2  = \langle \psi_q|r^2 |\psi_q\rangle/
\langle \psi_q|\psi_q\rangle$ with the pair wave function $\psi_q= 
g_{ql}\Delta
/(2E_{\bf q})$ can be evaluated  by using $r^2 = -\nabla_q ^2$.
In the weak coupling limit,
the zero-range analytic result of Ref. \cite{c} leads to $\xi^2= 0.5 \Delta
^{-2}(0) = 0.125 (\Delta U)^{-1}$.

\section{Numerical Results}

\vskip -5.cm
\postscript{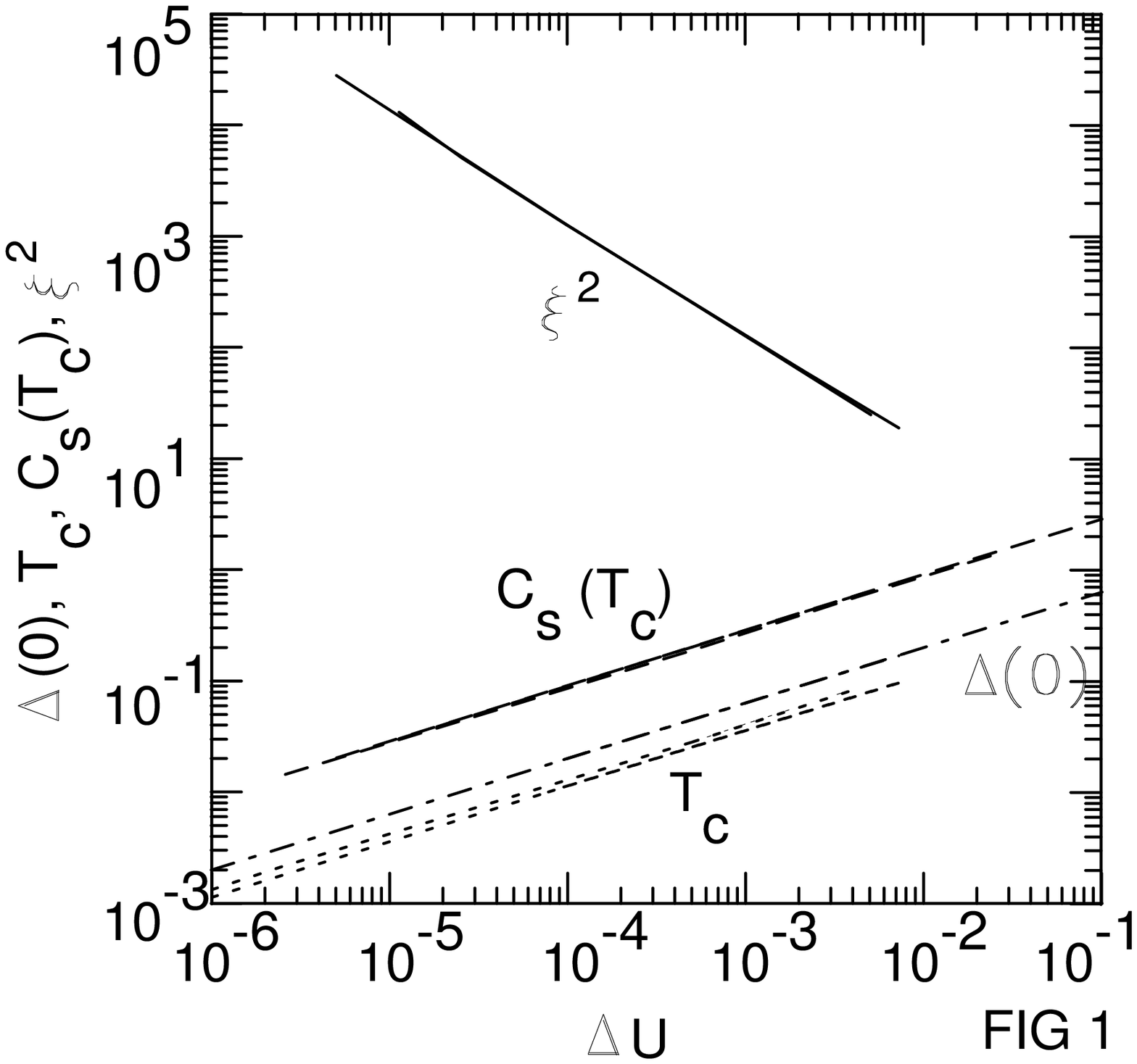}{0.8}    
\vskip -2.5cm
\noindent {\small {\bf Fig. 1} \ \  
{The plot of specific heat $C_s(T_c)$
 (dashed line),  $T_c$ (dotted line), gap parameter 
$\Delta(0)$ (dashed-dotted line) 
for $s$, $p$, and $d$ waves 
and $s$-wave pair radius $\xi^2$ (solid line)  versus zero-temperature
condensation energy per particle $\Delta U$ for different potential
parameters between $\alpha =1$ to $\infty$.  For the first two variables
there are two distinct lines; the upper one for $p$ and $d$ waves and the
lower one for $s$ wave. }}
\vskip 0.5cm
\vskip -0.33cm

We solve  coupled  equations
 (\ref{16}) and (\ref{15}) numerically using 
  dimensionless form factors $g_{ql}
=(\epsilon_q)^{l/2} [\alpha/(\epsilon_q+\alpha)]^{(l+1)/2} $ with
the correct
threshold behavior for small momenta,
where $\alpha $ is the range parameter.  Following Ref. \cite{t}
we calculate the dimensionless 
gap parameter  $\Delta (0)= g_{q(=1) l}\Delta$, $T_c$, $C_s(T_c)$, 
the $s$-wave pair radius $\xi^2$ at $T=0$ as well as $\Delta(T)$,
$\lambda(T)$, and $C(T)$ 
for different coupling.   In figure 1 we plot 
$\Delta (0)$, $T_c$, $C_s(T_c)$, and 
 $\xi^2$ at $T=0$ versus $\Delta U$ and find universal scalings. 
 The calculations were 
repeated for different potential ranges $\alpha$. We varied 
 $\alpha  $ from 1 to $\infty$ and found  figure 1  to 
be  insensitive to this variation in each partial wave. For $p$ and $d$ waves
  (\ref{16}) and (\ref{15}) diverge for $\alpha \to \infty$ and
calculations were performed for $\alpha = 1 $ to 10. The increase
in $\Delta U$ of figure 1 corresponds to an increase in coupling.  
We could
express this change  in coupling by  a change in  $B_2$
or in   Cooper pair binding  and plot the variables of figure 1 
in terms of these bindings  as in Ref. \cite{us}. Then each value of range
parameter leads to a distinct curve. However, if we express the variation in 
coupling by a variation of an observable of the superconductor, such as 
$\Delta U$ or $T_c$, universal 
potential-independent scalings are obtained. In each case the exponent and
the prefactor of each scaling relation are in excellent agreement with 
the analytic relation obtained above without form factors.

\vskip -5.cm
\postscript{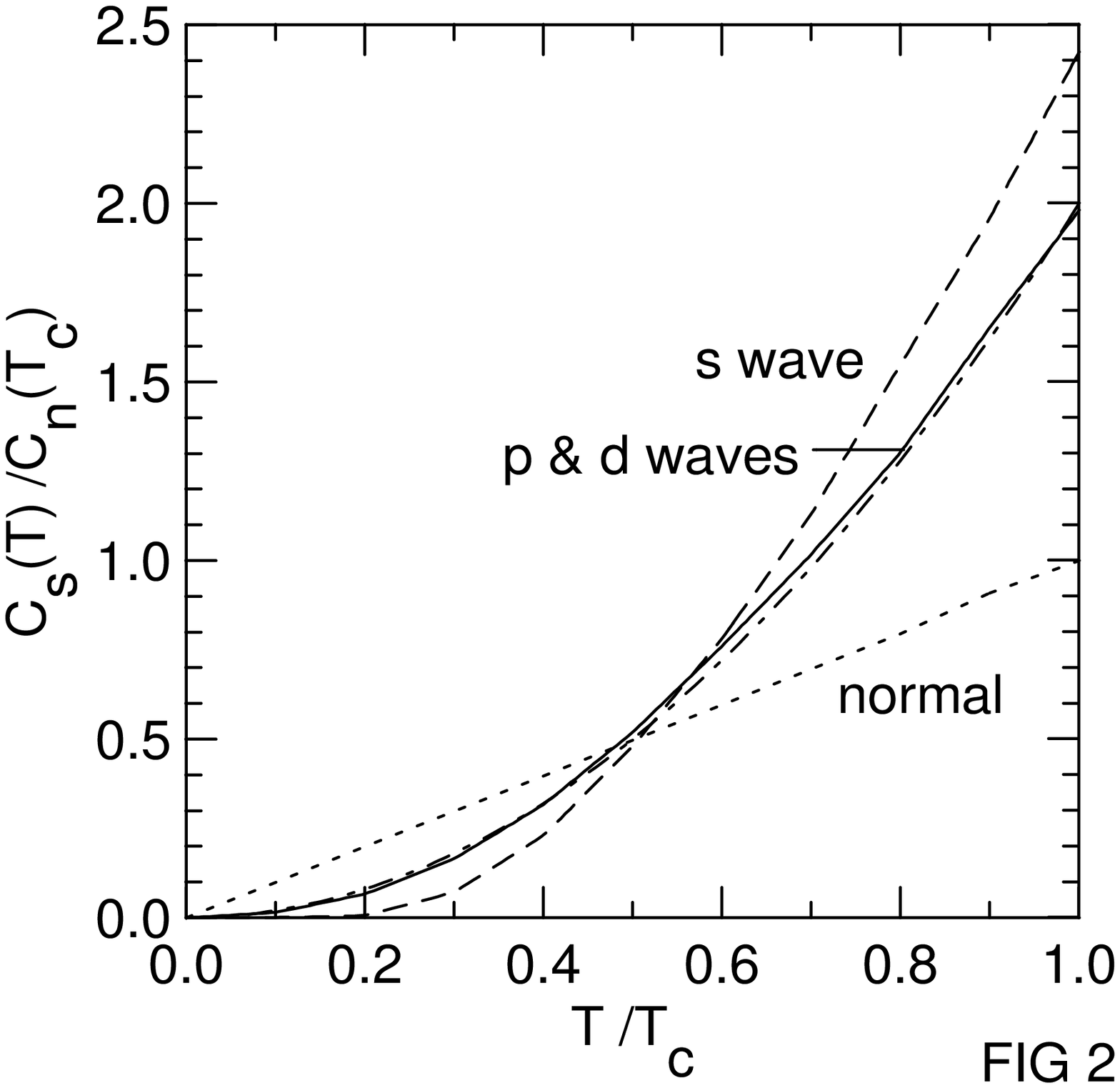}{0.8}    
\vskip -2.5cm
\noindent {\small {\bf Fig. 2} \ \  
{Specific heat $C_s(T)/C_n(T_c)$  versus $T/T_c$
 for $s$ (dashed line), $p$ and
$d$ (solid line) waves and potential parameters between $\alpha =1$ to
$\infty$.  The analytic fit (dashed-dotted line) 
$C_s(T)/C_n(T_c)=2(T/T_c)^2$,
to $p$ and $d$ waves is also shown. }}
\vskip 0.5cm
\vskip -0.33cm

The $T_c$ should not arbitrarily increase with coupling as figure 1 may 
imply. With increased coupling the  electron pairs should form
composite bosons which may undergo a phase transition under the action of a
residual interaction. According to a numerical study this transition happens 
at a temperature of 0.1 \cite{Dr}. This is why the
$T_c$ curve
in figure 1 has been plotted till about $T_c = 0.1$.
  For a very large class
of two-dimensional high-$T_c$ materials, the 
 $T_c $ 
has been estimated to be about 0.05 \cite{u}, which corresponds, for
$g_{ql}=1$, 
 to $B_2=0.004$. This small value of $B_2$ 
  is in the 
weak coupling  region where the universality of the present study should
hold. The  corresponding dimensionless 
pair radius ($\xi^2 \sim 80$)
at $T_c = 0.05$, as obtained from figure 1, is in  agreement with 
experimental analysis \cite{u}. 
Hence, the present study is relevant for these high-$T_c$ materials.

Next we studied the temperature dependence of
$\Delta(T)$,     $C_s(T)$, $\lambda(T)$ and $K(T)$ 
  for $T<T_c$.   For BCS
superconductors, these observables have exponential dependence 
on $T$ 
 as $T \to 0$ \cite{t,t2}. The two-dimensional high-$T_c$ superconductors have
a power-law dependence on $T$. In figures
 2, 3, and 4 we plot $C_s(T)/C_n(T_c)$,
 $\Delta\lambda(T)\equiv (\lambda(T)-\lambda(0))/\lambda(0)$, and 
 $K_s(T)/K_n(T)$ 
versus $T/T_c$, respectively. 
In these figures we
find  universal power-law dependence, essentially independent of potential 
range,   for $l\ne 0$. We find $C_s(T)\approx
2C_n(T_c)( T/T_c)^2$, $K_s(T)\approx
2K_n(T)( T/T_c)^{1.2}$  
for almost the entire temperature range and 
$\Delta\lambda(T)/\lambda(0) \sim (T/T_c)^{1.3}$ for small $T/T_c$. 
The $T^2$ dependence of
$C_s(T)$  was found in a theoretical study of
the Eliashberg equation \cite{t1} and in an analysis of experimental data
\cite{s}. The power-law dependence of  $\lambda(T)$ on $T$
was also conjectured before \cite{t2}. 
 The gap function $\Delta(T)$ has essentially the same universal 
behavior as in $s$ wave \cite{t}.  

\vskip -5.cm
\postscript{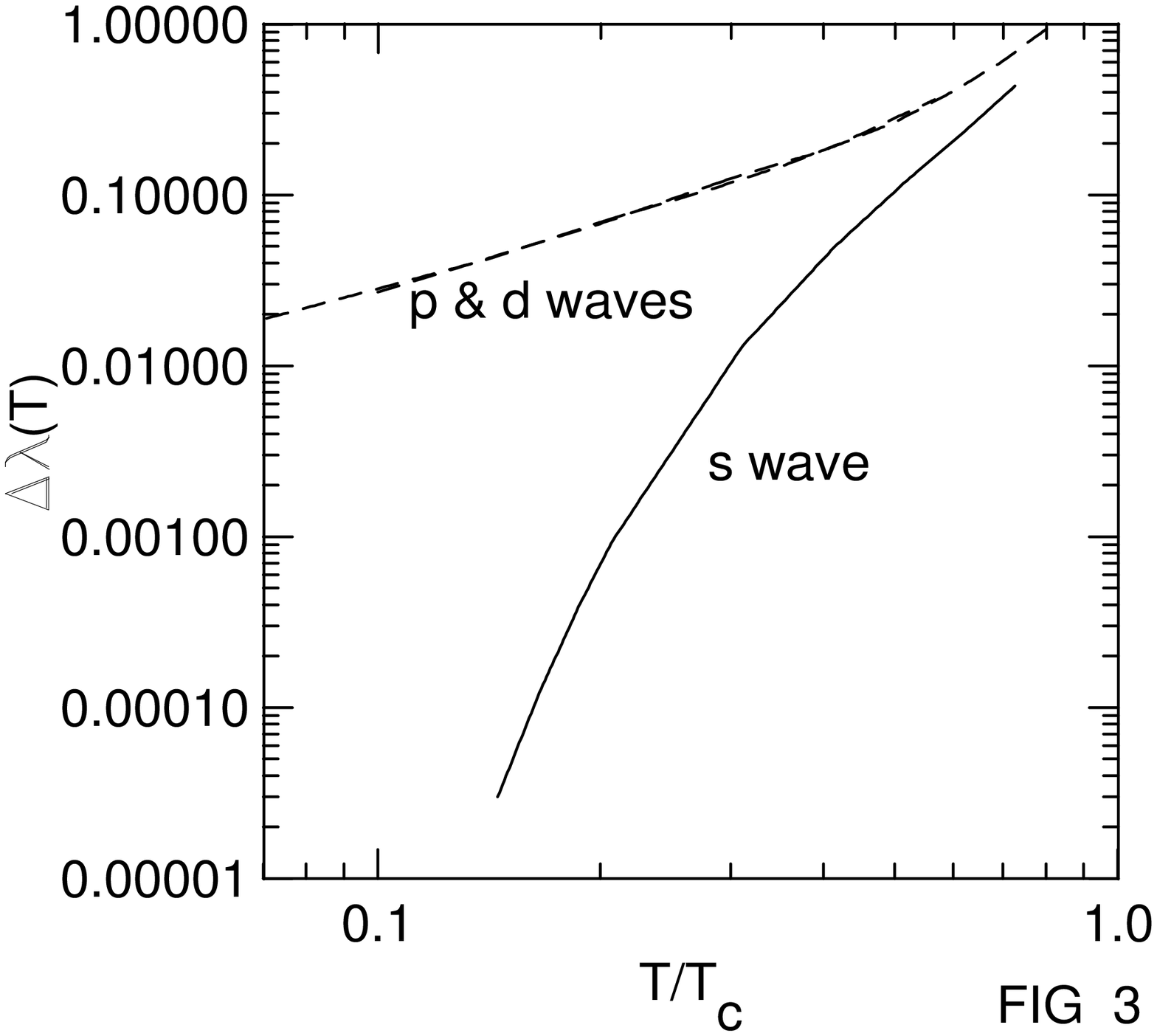}{0.8}    
\vskip -2.cm
\noindent {\small {\bf Fig. 3} \ \  
{Penetration depth $\Delta\lambda(T)$  versus $T/T_c$ for
$s$ (dashed line), $p$ and
$d$ (solid line) waves
and potential parameters between 
$\alpha =1$ to $\infty$.}}
\vskip 0.5cm
\vskip -0.33cm

\section{Summary}

Scalings are established among  $\Delta(0)$, $T_c$,
$C_s(T_c)$, and $\xi^2$, as function of $\Delta U$, independent of
potential range in $s$, $p$ and $d$  waves from a study of a renormalized 
 BCS
equation in two dimensions.  The present renormalized model yields a 
large $T_c$ and a small $\xi$ appropriate for some high-$T_c$ superconductors.
The $T$
dependence of $\Delta\lambda(T)$, $C_s(T)$ and $K_s(T)$
below $T_c$ in non-$s$ waves 
show  power-law
scalings distinct to some high-$T_c$ materials at low energies. No
power-law $T$ dependence is found in $s$ wave for 
$\Delta\lambda(T)$ and $C_s(T)$. Calculations performed with $\cos(l\theta)$
and $\sin(l\theta)$ angular dependences yielded identical results. 
Though we have used a separable potential, 
in view of the universal nature of the study we do not believe 
the present conclusions to be so peculiar as to have no general validity.
We have
repeated  the $s$ wave calculations with local Yukawa potential 
and found the results to be independent of potential in the weak-coupling
region. This is also in 
agreement with a suggestion by Leggett \cite{L}.
Although, there are controversies about a 
microscopic formulation of high-$T_c$ superconductors,
it seems that the two-dimensional 
$d$-wave BCS equation for weak  coupling  can be used to explain
some of their universal  scalings. A similar study of universality has
recently been performed in three dimensions employing the renormalized 
BCS equation \cite{th}, which also leads to an enhanced $T_c$ in the
weak-coupling limit.

\vskip -5.cm
\postscript{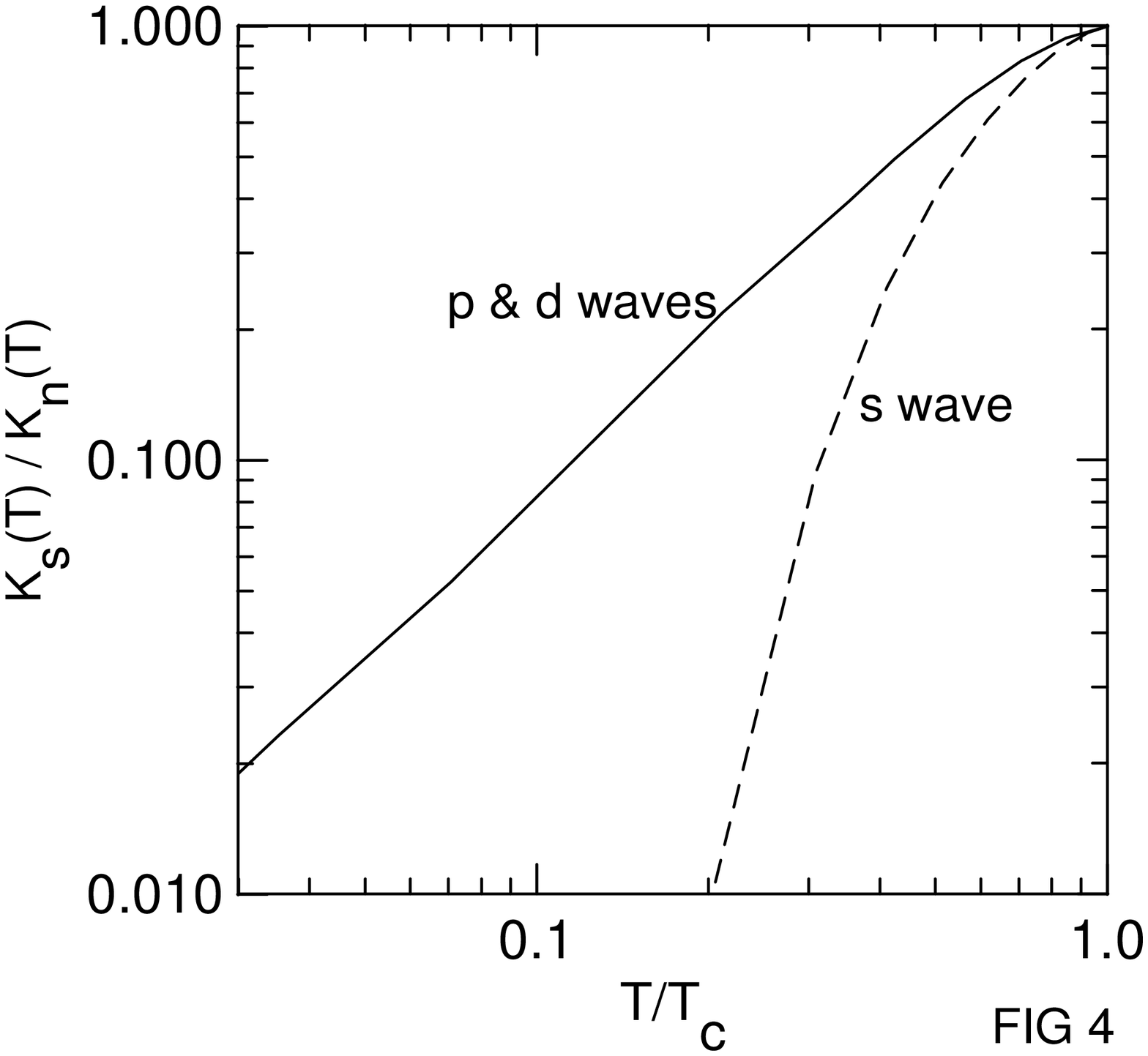}{0.8}    
\vskip -1cm
\noindent {\small {\bf Fig. 4} \ \  
{Thermal conductivity ratio $K_s(T)/K_n(T)$ versus $T/T_c$
 for $s$ (dashed line), $p$ and
$d$ (solid line) waves and potential parameters between $\alpha =1$ to
$\infty$.}}
\vskip 0.5cm

We thank Dr. M. Randeria, Dr. N. Trivedi  and Dr. M. de Llano
for helpful discussions and
Conselho Nacional de Desenvolvimento Cient\'{\i}fico e Tecnol\'ogico and 
Funda\c c\~ao de Amparo \`a Pesquisa do Estado de S\~ao Paulo
for 
financial support.

\end{document}